\documentclass[twocolumn,prl,showpacs]{revtex4}
\usepackage{epsf}

\newcommand{\fig}[1]	{Fig.~\ref{#1}}
\newcommand{\subbox}[1]	{{\mbox{\scriptsize #1}}}

\def\bi         {\begin{itemize}}
\def\ei         {\end{itemize}}
\def\benu	{\begin{enumerate}}
\def\eenu	{\end{enumerate}}
\def\bmat       {\left[ \begin{array}}
\def\emat       {\end{array} \right]}
\def\beq	{\begin{equation}}
\def\eeq	{\end{equation}}
\def\beqn       {\begin{eqnarray*}}
\def\eeqn       {\end{eqnarray*}}
\def\beqa       {\begin{eqnarray}}
\def\eeqa       {\end{eqnarray}}
\def\bquote	{\begin{quote}}
\def\equote	{\end{quote}}
\def\f          {\frac}
\def\bwide	{\begin{widetext}}
\def\ewide	{\end{widetext}}

\def\b          {\beta}

\def\m          {\mu}

\def\D          {\Delta}

\def\bk         {{\bf k}}

\def\bq         {{\bf q}}

\def\bv         {{\bf v}}

\begin{document}
\title{A novel route to a finite center-of-mass momentum pairing state; current driven FFLO state}
\author{Hyeonjin Doh}
\email{hdoh@physics.utoronto.ca}
%\affiliation{Department of Physics, University of Toronto, Toronto, 
%Ontario M5S 1A7, Canada}
\author{Matthew Song}
%\affiliation{Department of Engineering, University of Toronto, Toronto,
%\affiliation{Department of Physics, University of Toronto, Toronto, 
%Ontario M5S 1A7, Canada}
\author{Hae-Young Kee}
\email{hykee@physics.utoronto.ca}
\affiliation{Department of Physics, University of Toronto, Toronto, 
Ontario M5S 1A7, Canada}
\date{\today}
\begin{abstract}
The previously studied
Fulde-Ferrell-Larkin-Ovchinnikov (FFLO) state is stabilized by
a magnetic field via the Zeeman coupling 
in spin-singlet superconductors.
Here we suggest a novel route to achieve non-zero center-of-mass
momentum pairing states in superconductors with Fermi surface nesting.
We investigate two-dimensional superconductors under a uniform external current,
which leads to a finite pair-momentum of ${\bf q}_{e}$.
We find that an FFLO state with a spontaneous pair-momentum of ${\bf q}_{s}$ 
is stabilized above a certain critical current which depends on the direction
of the external current.
A finite ${\bf q}_s$ arises in order to make the total pair-momentum of 
${\bf q}_t(={\bf q}_s + {\bf q}_e)$ perpendicular to the nesting vector,
which lowers the free energy of the FFLO state,
as compared to the superconducting and normal states.
We also suggest experimental signatures of the FFLO state.
\end{abstract}

\maketitle

{\bf Introduction:} 
Fulde and Ferrell\cite{Fulde64pr}, and Larkin and Ovchinnikov\cite{Larkin65jetp}
predicted that Cooper pairs with  a nonzero center-of-mass momentum can
be stabilized under a finite magnetic field via the Zeeman coupling, 
when the Zeeman term dominates over the orbital effect. 
Recently, there has been growing interest in theoretical and 
experimental studies of the Fulde-Ferrell-Larkin-Ovchinnikov (FFLO) state in 
spin-singlet superconductors and trapped cold fermionic atoms 
with unequal densities\cite{Yang06}.
It has been suggested theoretically that the organic superconductors like
$\kappa$-(ET)$_2$ salts and $\lambda$-(ET)$_2$ salts,
heavy-fermion superconductors, and high $T_C$ cuprates are
promising candidates for observing the FFLO state.
The formation of a possible FFLO state has been very recently inferred
from  specific heat\cite{Bianchi03prl} and magnetization measurements
\cite{Martin05prb} on one of heavy fermion systems, 
CeCoIn$_5$ known to be a $d$-wave superconductor.

The main idea of the previously studied FFLO state is based on the well
known BCS theory that
Cooper pairs in spin-singlet superconductors are made of fermions with opposite spins.
When the magnetic field is acting on the spin of the electrons via the Zeeman coupling,
electrons tend to polarize along the direction of the magnetic field to gain
polarization energy, while the pairing of opposite spins  
is favorable for condensation energy. As a result of the competition,
the superconducting state 
undergoes a transition to the FFLO state, and the system 
eventually enters the normal state,
as the magnetic field is further increased.
It has been  shown that
the FFLO state can lead to an enhancement of the critical magnetic
field of up to 2.5 times the Pauli paramagnetic limit. 
It has also been suggested that
the FFLO state can be  realized when two species of fermion
have different densities in trapped cold fermionic atoms.\cite{Yang06}

\begin{figure}[hbt]
\epsfxsize=6cm
\epsffile{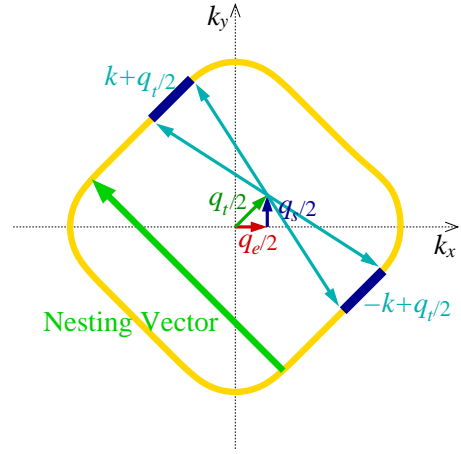}
\caption{
(Color online)
%A schematic view for Cooper-pair momenta, ${\bf q}_{e}$ and  ${\bf q}_{s}$.
The thick arrow indicates a nesting vector of a Fermi surface 
on a square lattice.
When an external current with a momentum $\bq_e$ 
is applied  along the $x$-axis,
a spontaneous pair-momentum of ${\bf q}_{s}$ is generated in such a way that
the total pair-momentum, ${\bf q}_t (={\bf q}_{e}+{\bf q}_{s})$ 
is perpendicular to the nesting vector. 
%The thin arrows denote the boundaries of
The region of Cooper pairing between electrons with
momentum ${\bf k}+{\bf q}_t/2$ and $-{\bf k}+{\bf q}_t/2$
is indicated by the dark bars . 
This pairing on the Fermi surface has the dominant contribution to lowering
the free energy of the FFLO state,
as compared to the superconducting and normal states.
\label{fig:FermiSurf}
}
\end{figure}

In this paper, we suggest a novel route to achieve the FFLO state.
We study two dimensional superconductors in the presence of uniform external
current, ${\bf j}_{e}$. When a current flows, Cooper pairs acquire
a finite external pair-momentum of ${\bf q}_{e}$, and
the superconducting state becomes unstable towards the normal state at a critical current.
If the underlying Fermi surface has nesting, we find that the strength of this critical
current depends on the direction of the current.
In the case of $s$-wave superconductors, the critical current is 
maximal when the  current is perpendicular to the nesting vector. 
%The external current produces the finite pair-momentum  ${\bf q}_e$ along the direction
%of the current. 
%This means that the stability of the superconducting state over normal
%state depends on the angle between the external current and the nesting vector.
We show that the FFLO state with a spontaneous pair-momentum of ${\bf q}_{s}$
is stabilized, when the current is not perpendicular to  the nesting vectors.
For example, a spontaneous pair-momentum, ${\bf q}_{s}$ is induced perpendicular to
the current,  when the current is along the $x$-axis of 
the square lattice as shown in \fig{fig:FermiSurf}.
The total pair-momentum of ${\bf q}_t$ (= ${\bf q}_{e} +{\bf q}_{s}$)
is determined to maximize the Cooper pairing between electrons
with momentum ${\bf k}+{\bf q}_t/2$ and $-{\bf k}+{\bf q}_t/2$ (in a lab frame).
This pairing on the Fermi surface has the dominant contribution to lowering
the free energy of the FFLO state, as compared to the superconducting and
normal states.

%The paper is organized as follow.
%Below we will discuss the effect of uniform current in superconductors,
%and present the BCS formalism to compute the free energy.
%Then we will show the temperature and current phase diagram for $s$- and $d$-wave
%superconductors, and finally discuss experimental signatures of the FFLO state.

{\bf Superconductors in the presence of a uniform supercurrent:} 
We consider a sample with thickness of $d \ll \xi$,
where $\xi$ is the coherence length. Under this condition, the magnitude
of the superconducting gap and the current are uniform across the sample.
When a uniform current flows, the Cooper-pair acquires a center-of-mass
momentum of ${\bf q}_{e}$. For an infinitesimal current 
%(or  ${\bf q}_e$), 
there will be no contribution from
the quasiparticles, and the supercurrent will be simply proportional
to $q_{e}$. However, the change of the quasiparticle spectrum in the presence of
a supercurrent known as a Doppler shift, is  written as follows.
\begin{equation}
E_{\bf k} =E_{\bf k}^0 +\bv_\bk \cdot \bq_e/2
\end{equation}
where $E_{\bf k}^0$ is the quasiparticle spectrum in the absence of the current,
%$( = \sqrt{\xi_{\bf k}^2+\Delta_{\bf k}^2})$ and
and
$\bv_\bk$ is the quasiparticle velocity.
%$\bv_{S,e} = q_{e}/(2m)$ for small $\bq_{e}$.
%${\bf v}_F (\phi)$
%$( =  \partial \xi_{\bf k} / \partial {\bf k})$
%is the quasiparticle velocity which depends on the angle $\phi$, around the 
%Fermi surface. 
The total supercurrent as a function of $\bq_{e}$ rises linearly at small
$\bq_{e}$ and reaches a maximum at a certain value of $\bq_{e}$. This value
determines the critical current at which 
the superconducting state become unstable towards the normal state.
Accordingly, 
a uniform supercurrent reduces the amplitude of the superconducting order parameter,
which has been extensively discussed in nodal superconductors.\cite{KeeHY04prb,Khavkine04prb1}

%We are seeking for a finite center-of-mass momentum pairing state (FFLO) as a stable
%state in the presence uniform currents.
One can intuitively understand that Cooper pairs with
a spontaneous center-of-mass momentum would
not occur, if the quasiparticle velocity is constant around the Fermi surface,
for $s$-wave superconductors  due to the Galilean invariance. 
However, once the quasiparticle velocity strongly 
depends on the angle around the Fermi surface  
due to Fermi surface nesting,
the strength of the critical current would vary as a function of the angle, $\phi$.
This anisotropy of the critical current is an essential ingredient for
an existence of the FFLO state.
In the case of anisotropic order parameters, such as an extended $d$-wave order parameter,
the anisotropic critical current occurs even for a circular Fermi surface. 
However, we found that the region of the FFLO state for this case is
too narrow to be practically realizable.

\begin{figure}
\epsfxsize=6cm
\epsffile{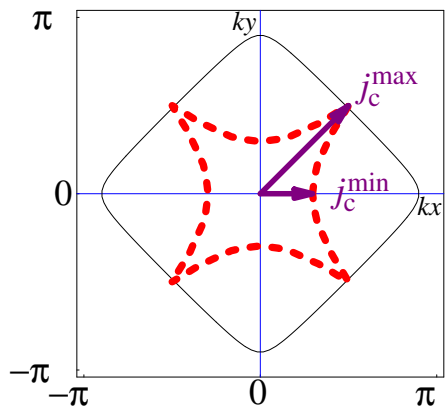}
\caption{
(Color online)
The direction dependence of the critical current value
(in arbitrary unit) is shown as the thick dashed line. 
The thin solid line is
the Fermi surface of the tight binding model on the square lattice.
The critical current has its maximum, ${j}_C^\subbox{max}$ when the current is
perpendicular to the nesting vector, 
while its minimum is ${j}_C^\subbox{min}$  along the $x$(or $y$)-direction.
\label{fig:Jc}
}
\end{figure}

In \fig{fig:Jc}, we show the direction dependence of the critical current for a $s$-wave superconductor
with the tight-binding electronic dispersion of the square lattice. 
The strength of the critical current strongly depends on its direction.
It attains its maximum, $j_C^\subbox{max}$, when the current is perpendicular to
the nesting vector.
When the current is applied along $x$-direction where
the critical current has its minimum, $j_C^\subbox{min}$, it is possible to develop 
a spontaneous pair-momentum of  ${\bf q}_{s}$ to extend the superconducting phase,
 even above $j_C^\subbox{min}$.
The development of the FFLO state leads to an enhancement of 
the critical current, similar to the enhancement of the critical magnetic field
in the previously studied FFLO phase. 
Below we carry out a free energy computation to find a possible FFLO
state and present temperature and current phase diagrams. 

{\bf Free energy and Phase diagram:} 
Here we adopt the standard BCS formalism to compute the free energy, and
determine the phase diagram as a function of temperature and external current.
The single particle Green  function $G (i\omega_n, {\bf k})$
in the presence of finite superflow is given by\cite{Maki69}
\beq
G^{-1}(i\omega_n,{\bf k}) =i \omega_n -
\bv_\bk
\cdot \bq_t/2 +
\xi_{\bf k} \rho_3 +\Delta({\bf k}) \rho_1 \sigma_1,
\eeq
where the single particle dispersion on the square lattice, 
$\xi_{\bf k} =  -t [\cos(k_x) + \cos(k_y) ] -\mu$.
The superconducting order parameter,
$\Delta({\bf k}) = \Delta_0 f({\bf k})$, where $f({\bf k})$ 
represents the relative momentum dependence of the order parameter.
$\rho_i$ and $\sigma_i$ are Pauli matrices in
particle-hole space and spin space, respectively.

The free energy is then given by
\beqa
F-\m N &=&
-\sum_\bk \f{\D_\bk^2}{E^0_\bk}
[f(E_{\bk +})+f(E_{\bk - })-1]
\nonumber\\
&-&\sum_{\bk\bk'}\f{V_{\bk\bk'}}{4}\f{\D_\bk\D_{\bk'}}
{E^0_{\bk}E^0_{\bk'}}
\left(1-f(E_{\bk + })-f(E_{\bk - })\right)
\nonumber\\&&\times
\left(1-f(E_{\bk' + })-f(E_{\bk' - })\right) \nonumber\\
&+& \sum_\bk \left[ (\xi_\bk -E^0_\bk) \right.
\nonumber\\&&\left. +\f{\ln (1+f(E_{\bk + }))(1+f(E_{\bk - }))}{\b} \right],
\eeqa
where $E_{\bk \pm} = E_{\bk}^0\pm\bv_\bk \cdot \bq_t/2 = \sqrt{\xi_k^2 +\Delta^2_k} \pm \bv_\bk \cdot \bq_t/2$. 
The total pair-momentum, ${\bf q}_t$ is
given by the sum of the external and spontaneous 
pair-momenta, ${\bf q}_t = {\bf q}_{e} + {\bf q}_{s}$.
The pairing interaction $V_{\bk\bk'}$ is assumed to have the form 
of $V_0 f{(\bk)} f{(\bk')}$. 

When an external current is applied along the parallel axes of
the square lattice, ${\bf q}_{e} =q_{e} {\hat x}$ (or ${\hat y}$),  
we find that the  minimum of the free energy as a function of the external current
shifts from ${\bf q}_{s}=0$ to a finite value of ${\bf q}_{s}$ above a critical value of
$q_{e}^{C1}$. 
The direction of ${\bf q}_{s}$ is perpendicular to the applied current in order to keep
constant input and output currents along the direction of the applied current.
The magnitude of ${\bf q}_{s}$ depends on the magnitude of the external current ${\bf q}_{e}$, 
and it is spontaneously arises so as to maximize the regions of electron
pairing as indicated by the dark bars in \fig{fig:FermiSurf}.

\begin{figure}
\epsfxsize=8cm
\epsffile{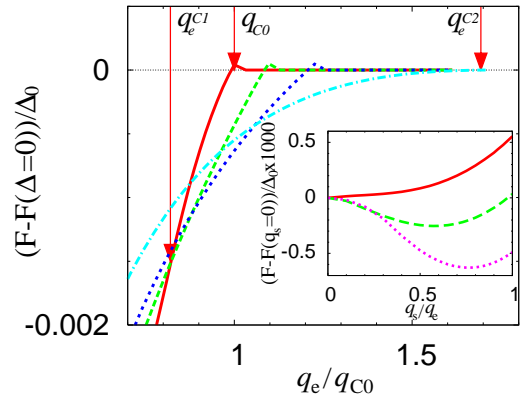}
\caption{
(Color online)
The free energy near zero temperature vs.  $q_{e}$
for various $q_{s}/q_{e}$. 
$q_{e}$ is normalized by $q_{C0}$ which is the critical current in the absence of the FFLO state.
The solid, dashed, dotted, and dot-dashed lines are for $q_{s}/q_{e} =0$,
0.4, 0.8 and 1.0, respectively.
Below $q^{C1}_{e} (=0.82 q_{C0})$, the uniform superconducting state
with $q_{s}=0$ has the lowest energy, while above $q^{C2}_{e} (=1.6935 q_{C0})$
the normal state has the lowest energy.
The FFLO states with finite values of 
$q_{s}$ are stabilized in the window of $ q^{C1}_{e} < q_{e} < q^{C2}_{e}$.
The inset shows the free energy vs. $q_{s}/q_{e}$ for various $q_{e}$ near the transition
between the uniform superconducting and the FFLO states.
The solid, dashed, dotted lines are for $q_{e}/q_{C0} = 0.8197$, 
0.8852, and  0.9836, respectively.
\label{fig:FreeEnergy}
}
\end{figure}

The result of the free energy near zero temperature as a function of the external current,
$q_{e}$ is shown in \fig{fig:FreeEnergy} for  several values of $q_{s}/q_{e}$ for the case of $s$-wave superconductors.
Note that above a critical value of $q^{C1}_{e}$, the solution with a finite ${q_{s}}$
has  lower energy than that of the uniform superconducting state.
It is also
important to note that the superconducting state becomes unstable towards
the normal state when $q_{e}$ exceeds the  critical value of $q^{C2}_{e}$, 
where the normal state has the lowest energy.
While our result of \fig{fig:FreeEnergy} is shown for a $s$-wave superconductor $f(\bk) =1$, 
the qualitative behavior of the phase transition to the FFLO state
in the presence of current 
does not depend on the  detailed nature  of the pairing symmetry, $f(\bk)$. 

\begin{figure}
\epsfxsize=8cm
\epsffile{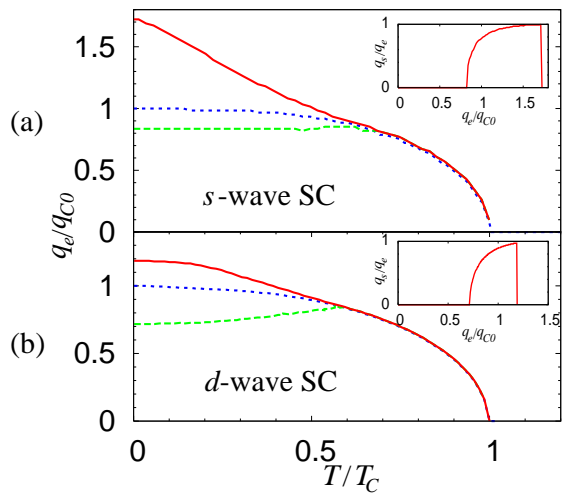}
\caption{
(Color online)
A temperature-current phase diagram for (a) $s$-wave and (b) $d_{x^2-y^2}$ superconductors.
The solid line is the second order phase boundary between the normal and FFLO states,
while the dashed line is the phase boundary 
between the uniform superconducting and FFLO states.
The dotted line indicates the phase boundary between the normal and superconducting
states in the absence of the FFLO state, where the critical current is denoted by
$q_{C0}$ at $T=0$. 
Temperature and external current are normalized by the critical temperature of $T_C$
and $q_{C0}$, respectively.
The inset shows the magnitude of the spontaneous pair-momentum $q_{s}$
normalized by $q_{e}$ vs. $q_{e}/q_{C0}$
% normalized by
%the critical pair momentum, $q_{c0}$ determined without the FFLO state.
\label{fig:PhaseDiagram}
}
\end{figure}

The temperature-current phase diagrams for  $s$- and $d_{x^2-y^2}$-wave superconductors are
shown in \fig{fig:PhaseDiagram}.
The solid line is the second order phase boundary separating the normal state
and the superconducting state. 
For $T/T_C < 0.7 $ (and 0.6 for a $d$-wave case), the FFLO state 
which is bounded by the dashed and solid lines is stabilized.
%in addition to an external pair-momentum, ${\bf q}_e$.
%Above $T/T_C > 0.7$, there is a second order transition between the superconducting
%and normal states.
The dotted line indicates the phase boundary between the normal and superconducting
states in the absence of the FFLO state, where the critical current at $T=0$
is denoted by $q_{C0}$. 
Note that the FFLO state enhances the critical current about 1.7 times for $\mu/(2t) =0.05$
at $T=0$ for a $s$-wave superconductor.
The transition between the FFLO and superconducting states is first order
for both $s$- and $d$-wave superconductors at low temperatures. 
%However, the nature of transition depends
%on a shape of the Fermi surface as well as a pairing symmetry, while the existence
%of the FFLO is independent of them.
The inset shows the magnitude of the spontaneous pair-momentum, $q_{s}$
normalized by the external pair-momentum $q_{e}$ as a function of $q_{e}/q_{C0}$.
% (normalized by
%the critical pair momentum, $q_{C0}$ determined without the FFLO state).
We have also analyzed the free energy for a p-wave superconductor with $f({\bk}) = \sin{k_x}
\pm i \sin{k_y}$, and found that the qualitative features of 
temperature-current phase diagram resemble those for the $s$- and $d$-wave cases
shown in \fig{fig:PhaseDiagram}.

{\bf Experimental signatures:}   
Here we assume that the superconducting gap and supercurrent will be uniform
across the system. This condition can be satisfied for a 
 sample with the thickness of $d \ll \xi$, where $\xi$ is the coherence length.
Below we offer a possible experimental set-up to detect the current driven FFLO state.

\begin{figure}
\epsfxsize=7cm
\epsffile{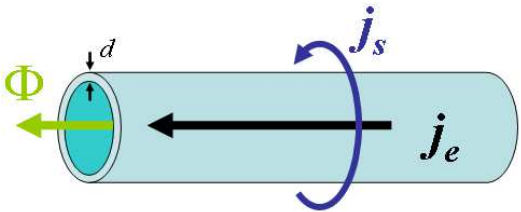}
\caption{
(Color online)
A schematic diagram for an experimental set-up. When an external current 
is applied along the axial direction of the tube, where the thickness
 $d \ll \xi$, a spontaneous current will be generated along the azimuthal
direction in the FFLO state. $\Phi$ denotes the flux induced by the spontaneous current. 
%where  (the system choose one of two directions of the current). 
%The spontaneous flux induced by the current can be ... for a circumference of 
%50 $nm$ for $d$-wave superconductors with the nested Fermi surface. 
\label{fig:ExperimentalSetup}
}
\end{figure}

Consider a superconductor of cylindrical geometry as shown in \fig{fig:ExperimentalSetup}.
If we apply the current along the axial direction, the spontaneous current, ${\bf j}_{s}$
will be generated along the azimuthal direction in
the FFLO state as shown in \fig{fig:ExperimentalSetup};
one of the two-fold degenerate directions of the spontaneous current will be chosen.
The induced current will generate a spontaneous flux, $\Phi$.
We find that a unit of flux can be generated when the circumference of the tube is
approximately 50 nm ($180 \times$lattice spacing) for $s$-wave superconductors
near half-filling with the tight-binding electronic dispersion when $q_{s}/q_{e} \approx 1$. 
%which can be
%realized in high T$_C$ cuprates.

The boundary conditions of an order parameter for  finite size samples
deserve some discussion.
We suggested the cylinder shape of the sample in the above experiment, so that
an order parameter with a finite current, 
$\Delta({\bf R}) \propto e^{i {\bf q}_{s} \cdot {\bf R}}$ 
can be stabilized, assuming that   
%In the above discussion, 
we can neglect  finite size effects on the length of the tube.
Depending on boundary, a linear combination of two plane waves,
such as $\Delta({\bf R}) \propto \cos{({\bf q}_{s} \cdot {\bf R})}$,
can be a possible solution,
which leads to a spatial modulation of the order parameter magnitude with
a periodicity of $2\pi/q_{s}$. 

Recently, the FFLO phase with unequal numbers of two hyperfine states of fermionic atoms
has been proposed in the context of cold fermionic atoms. 
The experiments have been carried out using Li atoms across a Feshbach resonance.
The experiment showed that  paired and unpaired fermions phase separate
when the imbalance is large\cite{Bedaque03}, 
but further experiments are needed to clarify the nature of the state 
for a small imbalance of densities\cite{Partridge06}.
We suggest that the current driven FFLO state can be also realized in
optical lattices
of cold atoms, where tight binding dispersions  have been found.

%\section{Summary:}
{\bf Summary:} 
We studied the  possible existence of the FFLO state in the presence
of an external current within BCS theory.
We found that the FFLO state with a spontaneous Cooper pair-momentum
can be stabilized 
in the presence of finite currents in superconductors with Fermi surface nesting.
The anisotropy of the critical current due to the nesting plays a crucial role 
for the existence of the FFLO state. A spontaneous pair-momentum is generated to
extend the superconducting phase and leads to
an enhancement of the critical current for certain directions.
The FFLO state can be found in anisotropic superconductors with a circular Fermi surface,
but the region of the FFLO state is too narrow to be practically realizable.
We suggested a possible experimental set-up to detect a spontaneous flux induced
by a spontaneous current, which will be a direct signature of the FFLO state.
We also discussed a possible realization of the FFLO state induced by a finite
current in optical lattices of cold fermionic atoms.

\begin{acknowledgments}
This work was supported by NSERC of Canada(HD, MS, HYK), Canada Research Chair,
Canadian Institute for Advanced Research, and Alfred P. Sloan
Research Fellowship(HYK).
\end{acknowledgments}

\bibliographystyle{apsrev}

\end{document}